\documentclass[aps,12pt,tightenlines,amsmath,amssymb,showpacs]{revtex4}
\usepackage{graphicx}
\usepackage{dcolumn}
\usepackage{bm}

\begin{document}

\title{PRIMORDIAL BLACK HOLES ARE AGAIN ON THE LIMELIGHT}

\author{Marco Roncadelli}
\affiliation{INFN, Sezione di Pavia, Via A. Bassi 6, I - 27100 Pavia, Italy}
\email{marco.roncadelli@pv.infn.it}

\author{Aldo Treves}
\affiliation{Physics Department, Universit\`a dell'Insubria, Via Valleggio 11, I - 22100 Como, Italy, and associated to INFN and INAF}
\email{aldo.treves@uninsubria.it}

\author{Roberto Turolla}
\affiliation{Physics Department, Universit\`a di Padova, Via Marzolo 8, I - 35131 Padova, Italy, associated to INFN and INAF and Mullard Space Science Laboratory, University College London, Holbury St. Mary, Dorking, Surrey, RH5 6NT, United Kingdom}
\email{roberto.turolla@pd.infn.it}

\date{\today}
\begin{abstract} 
We derive a strong upper bound on the amount of Primordial Black Holes (PBHs) that can still be present in the Universe. Gravitational capture of PBHs by the Milky Way stars during their formation and subsequent accretion would produce a dramatic depletion of disk stars and especially of white dwarfs, unless the average cosmic density and mass of PBHs are severely constrained. Our finding also helps to discriminate among the various production mechanisms of PBHs. Moreover, we show that a star becomes overluminous  before its disappearance into a PBH for a time span independent of its mass, thereby providing a characteristic observational signature of the considered scenario. We stress that our result allows for the existence of stellar-mass black holes in a mass range that is forbidden by standard stellar evolution.
\end{abstract}

\maketitle

\bigskip

\normalsize\baselineskip=15pt

The formation of black holes in the early Universe -- called Primordial Black Holes (PBHs) -- is regarded as an inescapable implication of the 
standard cosmological model. As recognized long ago, PBHs should be produced by the collapse of overdense regions when the Hubble radius equals 
their Schwarzschild radius~\cite{zh}. In addition, several other mechanisms are expected to trigger PBH formation, like the various phase-transitions 
occurring during the cosmic evolution, the collisions of cosmic loops as well as of domain walls, the inflationary reheating and a temporary 
softening of the cosmic equation of state~\cite{carr1}. Moreover, their existence turns out to be a general feature of brane cosmology 
arising in theories with compactified extra dimensions~\cite{brane}. 

A most striking feature of PBHs is the phenomenon of Hawking radiation~\cite{hawrad}, which leads to their evaporation on a characteristic time scale 
$t_{\rm ev} \simeq 10^{10} (M_{\rm PBH}/10^{15} \, {\rm g})^3 \, {\rm yr}$. Hence, PBHs with mass $M_{\rm PBH} < 10^{15} \, {\rm g}$ should 
have already evaporated and the associated emission could have affected the cosmic evolution in several respects. Extensive work has 
addressed PBHs close to the evaporation limit $M_{\rm PBH} \simeq10^{15} \, {\rm g}$, since in this case the Hawking radiation may produce 
observable effects in the present Universe~\cite{natrev}. So far, no positive evidence for PBHs has been reported and this circumstance allows 
to set upper bounds on their present density. Actually, the strongest one comes from EGRET observations of the gamma-ray background in the 
energy range $30 \, {\rm MeV} < E < 120 \, {\rm GeV}$ and implies that the contribution to the cosmic density parameter $\Omega$ from PBHs 
with mass close to the evaporation limit is ${\Omega}_{\rm PBH} < 10^{- 8}$~\cite{boundgamma}. We stress that no similar constraint is 
available for PBHs with $M_{\rm PBH} > 10^{15} \, {\rm g}$ because they are still present today and their Hawking emission is negligible. 

As a matter of fact, PBHs with mass $M_{\rm PBH} > 10^{15} \, {\rm g}$ have attracted specific attention in the past. Since they behave as dynamically 
cold objects, they are very good candidates for the Cold Dark Matter (CDM) that dominates the mass budget of the Universe~\cite{carr2}. For instance, 
it has been argued that PBHs in the mass range $5 \cdot 10^{15} \, {\rm g} < M_{\rm PBH} < 10^{21} \, {\rm g}$ can be produced by a primordial 
spectrum (with a characteristic mass scale) of inflationary origin and can make up a significant fraction of CDM~\cite{kiefer}. A more radical option is that 
PBHs could be the dominant form of CDM in galaxies, thereby triggering their formation and so dispensing with the need of new elementary particles as CDM constituents~\cite{spergel}. Finally, the proposal has been put foreward that PBHs might ultimately produce the supermassive black holes residing in the galactic nuclei~\cite{supermassive}.

Surprisingly enough, little attention has been devoted so far to the astrophysical effects of PBHs with mass $M_{\rm PBH} > 10^{15} \, {\rm g}$, most probably on the assumption that they hardly interact with the surrounding low-density matter and in particular that they undergo practically no accretion from the interstellar medium. Yet, if PBHs are present in sufficiently large number, gravitational capture by stars in the Milky Way disk during their formation process becomes unavoidable~\cite{bambi,abram}. Our aim is to explore the astrophysical implications of this phenomenon. Our conclusions set a strong constraint on the amount of PBHs in the present Universe, while leaving room for the existence of stellar-mass black holes in a mass range that is forbidden by standard stellar evolution.

We start by remarking that much in the same way as it happens for nonbaryonic CDM which is currently thought to make up the galactic halos, 
PBHs of any mass above the evaporation limit are expected to undergo the same collapse process just because they are dynamically cold objects. We recall that in first approximation galactic dark halos are currently described by the Cored Spherical Isothermal (CSI) model, with density profile 
$\rho (r) = a^2 / (a^2 + r^2)$ ($a$ is the core radius) and Maxwellian velocity distribution with constant one-dimensional dispersion $\sigma$~\cite{binney}. 
Note that $\sigma$ is independent of the mass of any halo constituent because the equilibrium distribution is achieved through violent relaxation~\cite{binney}. So, it looks natural to {\rm assume} that the distribution of PBHs inside the Milky Way should be described by the CSI model as well, which dictates
\begin{equation}
\rho_{\rm CDM} (R,z) = \rho_{\rm CDM} (R_0,0) \, \left( \frac{a^2 + R_0^2}{a^2 + R^2 + z^2} \right)~,     
\label{a1a2}
\end{equation}
where cylindrical Galactocentric coordinates are used  and $R_0 \simeq 8 \, {\rm kpc}$ denotes our Galactocentric distance. 
Moreover, we expect the overdensity of PBHs in the Milky Way (as compared to their average cosmic density) to roughly equal that of nonbaryonic 
CDM, thereby implying $n_{\rm PBH} (R,z) \, M_{\rm PBH} \simeq \rho_{\rm CDM} (R,z) \, \left( \Omega_{\rm PBH}/ \Omega_{\rm CDM} \right)$, where we are assuming that PBHs have a characteristic mass $M_{\rm PBH}$. Various studies yield $a \simeq 6.4 \, {\rm kpc}$, ${\rho}_{\rm CDM} (R_0,0) \simeq 0.3 \, ({\rm GeV} /c^2) \, {\rm cm}^{- 3}$ and $\sigma \simeq 1.6 \cdot 10^7 \, {\rm cm} \, {\rm s}^{- 1}$ as preferred values~\cite{shm}. In addition, the CDM contribution to $\Omega$ is ${\Omega}_{\rm CDM} \simeq 0.3$. Hence, we get 
\begin{equation}
n_{\rm PBH}{(R,z)} \simeq 1.8 \cdot 10^{- 39} \left(\frac{10^{15} \, {\rm g}}{M_{\rm PBH}} \right)  \left(\frac{a^2 + R_0^2}{a^2 + R^2 + z^2} \right)
{\Omega}_{\rm PBH} \ {\rm cm}^{-3}~,
\label{a1}
\end{equation}
and so the distribution function of PBHs in the Milky Way is
\begin{equation}
f_{\rm PBH} (R,z,v) = \frac{n_{\rm PBH}(R,z)}{(2 \pi \sigma^2)^{3/2}} \, e^{- v^2 /2 \sigma^2}~.
\label{a1a3}
\end{equation}

Next, let us go back to the epoch when disk stars formed in the Milky Way -- which occurred a few Gigayears ago -- and let us focus our attention on a single protostar cloud located at $(R,0)$, i.e. a Giant Molecular Cloud (GMC) of typical mass $M_{\rm GMC} \simeq 10^{39} \, {\rm g} $ and linear size $R_{\rm GMC} \simeq 10 \, {\rm pc}$.  According to current wisdom, it produces many stars upon fragmentation associated with the isothermal gravitational collapse. Therefore, the number of {\it unbound} PBHs initially contained in such a GMC is simply its volume times the number density given by Eq. (\ref{a1}), namely
\begin{equation}
N_{\rm PBH,i}^{\rm GMC}{(R,0)} \simeq  2.2 \cdot 10^{20} \left(\frac{10^{15} \, {\rm g}}{M_{\rm PBH}} \right)  \left(\frac{a^2 + R_0^2}{a^2 + R^2} \right)
{\Omega}_{\rm PBH}~.
\label{a1A}
\end{equation}
Owing to the GMC collapse, a PBH inside it feels a time-decreasing gravitational potential. So, its energy gets lowered, thereby allowing a fraction of 
PBHs  in the collapsing GMC to get {\it captured}~\cite{precis}. Below, we proceed to a quantitative estimate of this effect, following closely the analysis of Steigman, Sarazin, Quintana and Faulkner (SSQF)~\cite{steigman}. As stressed by these authors, the realistic situation is the one in which the collapse time $t_{\rm c} \equiv \Phi_{\rm GMC} /(\partial \Phi_{\rm GMC} / \partial t)$ is much longer than the free-fall time $t_{\rm ff} \equiv (R_{\rm GMC}^3/ G M_{\rm GMC})^{1/2}$, where $\Phi_{\rm GMC}$ denotes the cloud gravitational potential. Assuming that the PBH distribution function in the considered GMC is given by Eq. (\ref{a1a3}), by going through the same steps of SSQF we find that the total number of PBHs captured in the course of the GMC collapse 
is~\cite{precis2}
\begin{equation}
N_{\rm PBH,c}^{\rm GMC}{(R,0)} \simeq  1.1 \left(\frac{G M_{\rm GMC}}{R_{\rm GMC} \, {\sigma}^2} \right)^{3/2} N_{\rm PBH,i}^{\rm GMC}{(R,0)}~, 
\label{a1B}
\end{equation}
that is
\begin{equation}
N_{\rm PBH,c}^{\rm GMC}{(R,0)} \simeq 1.9 \cdot 10^{17} \left(\frac{10^{15} \, {\rm g}}{M_{\rm PBH}} \right)  \left(\frac{a^2 + R_0^2}{a^2 + R^2} \right)
{\Omega}_{\rm PBH}~.
\label{a1C}
\end{equation}
Since even the isothermal collapse has an efficiency of at most $40 \, \%$~\cite{scalo}, we expect that no more than $10^5$ stars should form out of a single GMC. As a consequence, the average number of PBH gravitationally bound to each such star should be
\begin{equation}
N_{\rm PBH,c}^{*}{(R,0)} \simeq 1.9 \cdot 10^{12} \delta \left(\frac{10^{15} \, {\rm g}}{M_{\rm PBH}} \right)  \left(\frac{a^2 + R_0^2}{a^2 + R^2} \right)
{\Omega}_{\rm PBH}~,
\label{a1D}
\end{equation}
where the dilution factor $\delta$ accounts for the possibility that some PBHs end up in the intracluster gas rather than in a star. Thus, we see that on average every star in the Milky Way disk contains at least one PBH provided that
\begin{equation}
M_{\rm PBH} < 1.9 \cdot 10^{27} \delta  \left(\frac{a^2 + R_0^2}{a^2 + R^2} \right) {\Omega}_{\rm PBH} \, {\rm g}~,
\label{a1E}
\end{equation}
where now $R$ stands for the star's Galactocentrc distance.

Once a PBH with mass $M_{\rm PBH}$ gets captured by a star, it sinks towards the center and starts accreting. In such a situation, its mass increases with time and will be denoted by $M_{\rm PBH}(t)$. Given that PBHs are produced in the early Universe and since we are dealing with a stellar population formed a few Gigayears ago, we can assume $M_{\rm PBH} \simeq M_{\rm PBH}(0)$, while the present value of a captured PBH mass is $M_{\rm PBH}(t_{\rm H})$, where $t_{\rm H} \simeq1.3 \cdot 10^{10} \, {\rm yr}$ is the Hubble time. Similarly, the resulting accretion luminosity $L_{\rm acc}(t)$ increases, eventually becoming comparable to the original stellar luminosity $L_*$ -- say, $L_{\rm acc} \simeq 0.1 \, L_*$ -- on a characteristic time scale $t_{\rm lum}$. It may also happen that the whole star gets swallowed by the PBH on a characteristic time $t_{\rm sw}$. Below, this picture will be worked out in a quantitative fashion.

As a preliminary step, we recall a few basic facts that are instrumental to our subsequent analysis. The PBH Schwarzschild radius is $R_s \simeq 1.5 \cdot 10^{- 13} (M_{\rm PBH}/ 10^{15} \, {\rm g}) \, {\rm cm}$, its Hawking temperature is $T_H \simeq 1.2 \cdot 10^{11} (10^{15} \, {\rm g}/ M_{\rm PBH}) \, {\rm K} $ and the corresponding luminosity is $L_{H} \simeq 3.3 \cdot 10^{15} \, (10^{15} \, {\rm g} / M_{\rm PBH})^2 \, {\rm erg} \, {\rm s}^{-1}$. Although $T_H$ can largely exceed that of stellar matter, $L_{H}$ is totally negligible in comparison to $L_*$. Within the Bondi accretion theory~\cite{accretion}, the accretion radius of any object of mass $M$ is $R_{\rm acc} \simeq 1.5 \cdot 10^{-4} \left(M /10^{15} \, {\rm g} \right) \left(10^4 \, {\rm K} / T_{\rm bd} \right) \, {\rm cm}$, while the accretion rate is given by
\begin{equation}
\frac{dM}{dt} \simeq 9.5 \cdot 10^{4} \left(\frac{M}{10^{15} \, {\rm g}} \right)^2 \, \left(\frac{{\rho}_{\rm bd} }{{\rm g} \, {\rm cm}^{-3}} \right)  \left(\frac{10^4 \, {\rm K}}{T_{\rm bd}} \right)^{3/2} \ {\rm g} \, {\rm yr}^{-1}~,
\label{a5}
\end{equation}
where $\rho_{\rm bd}$ and $T_{\rm bd}$ denote the density and the temperature, respectively, of matter at $R_{\rm acc}$. Eq. (\ref{a5}) can be trivially solved to yield
\begin{equation}
\frac{10^{15} \, {\rm g}}{M (t)} = \frac{10^{15} \, {\rm g}}{M (0)} - 9.5 \cdot 10^{- 11} \left(\frac{{\rho}_{\rm bd} }{{\rm g} \, {\rm cm}^{-3}} \right)  
\left(\frac{10^4 \, {\rm K}}{T_{\rm bd}} \right)^{3/2} \left(\frac{t}{{\rm yr}} \right)~.
\label{a5bis1}
\end{equation}
The associated accretion luminosity is
\begin{equation}
L_{\rm acc} \simeq \epsilon \, c^2 \, \frac{dM}{dt} ~,
\label{a7}
\end{equation}
where $\epsilon$ denotes the efficiency of the accretion process. The issue of the efficiency in spherical accretion has been widely debated in the past and $\epsilon$ was found to depend on the accretion rate~\cite{ntz}. In the regime of large optical depth -- typical of the stellar interior -- it turns out that $10^{-5} \lesssim \epsilon \lesssim 10^{-4}$. Whenever necessary, we will conservatively take $\epsilon \simeq 10^{-5}$. Generally speaking, as the accretion proceeds 
$L_{\rm acc}$ increases until the Eddington luminosity 
\begin{equation}
L_{\rm E} \simeq 6.5 \cdot 10^{19} \left(\frac{M}{10^{15} \, {\rm g}} \right) \ {\rm erg} \, {\rm s}^{-1}
\label{a7WW}
\end{equation}
is attained, after which the accretion process gets Eddington-limited and the accretion rate becomes
\begin{equation}
\frac{dM}{dt} \simeq 2.3 \cdot 10^6 \, {\epsilon}^{- 1} \left(\frac{M}{10^{15} \, {\rm g}} \right) \ {\rm g} \, {\rm yr}^{-1}~.
\label{a5z}
\end{equation}
It proves convenient to represent the solution of Eq. (\ref{a5z}) in the form
\begin{equation}
M (t) = M(0)\, 10^{X(t)}~,
\label{a7c1}
\end{equation} 
with $X(t) \simeq 1 \cdot 10^{-9} \,{\epsilon}^{-1} \, \left(t / {\rm yr} \right)$.

Coming back to our main line of development, we inquire about the fate of a disk star that has indeed captured a PBH~\cite{clayton}. Specifically, we will address both the case of a Sun-like star and that of such a star having evolved into a white dwarf.

As far as the behaviour of a Sun-like star is concerned, we assume for definiteness $M_* \simeq 2 \cdot 10^{33} \, {\rm g}$, $L_* \simeq 4 \cdot 10^{33} \, 
{\rm erg} \, {\rm s}^{-1}$, ${\rho}_{bd} \simeq 10 \, {\rm g} \, {\rm cm}^{-3}$ and $T_{bd} \simeq 10^7 \, {\rm K}$. Then the PBH settles at the centre because of dynamical friction on a time scale $t_{\rm df}^{\rm sl} \simeq 7.4 \cdot 10^9 \left( v_{\rm PBH} / 10^7 \, {\rm cm} \, {\rm s}^{-1} \right)^3 (10^{15} {\rm g}/ M_{\rm PBH}) \, {\rm yr}$~\cite{binney}, the accretion radius is $R_{\rm acc} \simeq 1.5 \cdot 10^{-7} \left( M_{\rm PBH} / 10^{15} \, {\rm g} \right) \, {\rm cm}$~\cite{comment2} and the accretion luminosity turns out to be
\begin{equation}
L_{\rm acc}^{\rm sl} \simeq 8.6 \cdot 10^{14} \, \epsilon \left(\frac{M_{\rm PBH}}{10^{15} \, {\rm g}} \right)^2 \ {\rm erg} \, {\rm s}^{-1}~.
\label{a7}
\end{equation} 
The comparison of $L_{\rm acc}^{\rm sl}(t)$ with $L_{\rm E}(t)$ shows that three possibilities can be realized. (1) Accretion is never Eddington-limited: this happens for $L_{\rm acc}^{\rm sl}(t_{\rm H}) < L_{\rm E}(t_{\rm H})$,  namely for $M_{\rm PBH}(t_{\rm H}) < 7.6 \cdot 10^{19} \, {\epsilon}^{- 1} \, {\rm g}$, i.e. $M_{\rm PBH}(t_{\rm H}) < 7.6 \cdot 10^{24} \, {\rm g}$. Owing to Eq. (\ref{a5bis1}), such a condition translates into $M_{\rm PBH} < 2.6 \cdot 10^{18} \, {\rm g}$ (regardless of $\epsilon$). Correspondingly, the presence of the PBH inside the star is totally harmless, since $L_{\rm acc}^{\rm sl}(t_{\rm H}) \ll L_*$ and $M_{\rm PBH}(t_{\rm H}) \ll M_*$. (2) Accretion is always Eddington-limited: this takes place for $L_{\rm acc}^{\rm sl}(0) > L_{\rm E}(0)$,  namely for $M_{\rm PBH} > 7.6 \cdot 10^{19} \, {\epsilon}^{- 1} \, {\rm g}$, i.e. $M_{\rm PBH} > 7.6 \cdot 10^{24} \, {\rm g}$. Clearly, $t_{\rm sw}$ is fixed by the condition $M_{\rm PBH}(t_{\rm sw}) \simeq M_*$. Hence, by setting $M (0) = M_{\rm PBH} \equiv 10^Y \, {\rm g}$ and $M (t) = M_*$, Eq. (\ref{a7c1}) entails $t_{\rm sw} \simeq (33.3 - Y) \cdot 10^9 \, {\epsilon} \, {\rm yr}$, i.e. $t_{\rm sw} \simeq (33.3 - Y) \cdot 10^4 \, {\rm yr}$. Since now $Y > 19.9 - {\rm log} \, {\epsilon}$, we deduce $t_{\rm sw} < (13.4 + {\rm log} \, {\epsilon}) \cdot 10^9 \, {\epsilon} \, {\rm yr}$, i.e. $t_{\rm sw} < 8.4 \cdot 10^4 \, {\rm yr}$. Moreover, $t_{\rm lum}$ is fixed by the requirement $L_{\rm E}(t_{\rm lum}) \simeq 0.1 \, L_*$. Accordingly, Eq. (\ref{a7WW}) yields $M_{\rm PBH}(t_{\rm lum}) \simeq 6.2 \cdot 10^{27} \, {\rm g}$ and using again Eq. (\ref{a7c1}) we find $t_{\rm lum} \simeq (27.8 - Y) \cdot 10^9 \, {\epsilon} \, {\rm yr}$, i.e. $t_{\rm lum} \simeq (27.8 - Y) \cdot 10^4 \, {\rm yr}$. So, for a time span of about $5.5 \cdot 10^9 \, {\epsilon} \, {\rm yr}$, i.e. $5.5 \cdot 10^4 \, {\rm yr}$ the star becomes overluminous before disappearing (regardless of $M_*$). (3) Accretion starts in the Bondi regime and next gets Eddington-limited for $2.6 \cdot 10^{18} \, {\rm g} < M_{\rm PBH} <  7.6 \cdot 10^{19} \, {\epsilon}^{- 1} \, {\rm g}$, i.e. $2.6 \cdot 10^{18} \, {\rm g} < M_{\rm PBH} <  7.6 \cdot 10^{24} \, {\rm g}$. The turnover occurs at time $t_{\rm to}$ such that $M_{\rm PBH}(t_{\rm to}) \simeq 7.6 \cdot 10^{19} \, {\epsilon}^{- 1} \, {\rm g}$ and can be found from Eq. (\ref{a5bis1}) by setting $M (0) = M_{\rm PBH}$ and $M (t) = M_{\rm PBH}(t_{\rm to})$. We obtain $t_{\rm to} \simeq 3.3 \cdot 10^{13} \left[ \left(10^{15} \, {\rm g}/ M_{\rm PBH} \right) - 1.3 \cdot 10^{-5} \, {\epsilon} \right] {\rm yr}$ and hence we have $t_{\rm to} < 1.3 \cdot 10^{10} \, {\rm yr}$. Accretion next gets Eddington-limited and proceeds much more rapidly, as described above. 

We next carry out a similar investigation for a white dwarf, taking for definiteness $M_* \simeq 2 \cdot 10^{33} \, {\rm g}$, $L_* \simeq 7 \cdot 10^{30} \, {\rm erg} \, {\rm s}^{-1}$, ${\rho}_{bd} \simeq 10^6 \, {\rm g} \, {\rm cm}^{-3}$ and $T_{bd} \simeq 10^7 \, {\rm K}$. Correspondingly, $R_{\rm acc}$ is still given by the previous expression~\cite{comment2}, whereas now the PBH settles at the centre because of dynamical friction on a time scale $t_{\rm df}^{\rm wd} \simeq 10^{-5} \, t_{\rm df}^{\rm sl}$ and the accretion luminosity becomes $L_{\rm acc}^{\rm wd} \simeq 10^5 \, L_{\rm acc}^{\rm sl}$. Proceeding as above -- namely comparing $L_{\rm acc}^{\rm wd}(t)$ with $L_{\rm E}(t)$ -- the following scenario emerges. (1) At variance with the previous case, a pure Bondi accretion regime is ruled out, because now  ${\rho}_{bd}$ is larger by a factor $10^5$ and so it would demand $M_{\rm PBH} < 2.6 \cdot 10^{13} \, {\rm g}$, 
which is against our main assumption. (2) The same argument implies that accretion is always Eddington-limited for $M_{\rm PBH} > 7.6 \cdot 10^{14} \, {\epsilon}^{- 1} \, {\rm g}$, i.e. $M_{\rm PBH} > 7.6 \cdot 10^{19} \, {\rm g}$. Since $M_*$ is unchanged, $t_{\rm sw}$ is the same as before. However, we presently have $Y > 14.9 - {\rm log} \, {\epsilon}$, which entails $t_{\rm sw} < (18.4 + {\rm log} \, {\epsilon}) \cdot 10^9 \, {\epsilon} \, {\rm yr}$, i.e. $t_{\rm sw} < 1.4 \cdot 10^5 \, {\rm yr}$. Finally, we evaluate $t_{\rm lum}$ from the condition $L_{\rm E}(t_{\rm lum}) \simeq 0.1 \, L_*$. Correspondingly, Eq. (\ref{a7WW}) gives $M_{\rm PBH}(t_{\rm lum}) \simeq 1.1 \cdot 10^{25} \, {\rm g}$ and from Eq. (\ref{a7c1}) we get $t_{\rm lum} \simeq (25 - Y) \cdot 10^9 \, {\epsilon} \, {\rm yr}$, i.e. $t_{\rm lum} \simeq (25 - Y) \cdot 10^4 \, {\rm yr}$. Hence, also a white dwarf becomes overluminous for a time span of about $8.3 \cdot 10^9 \, {\epsilon} \, {\rm yr}$, i.e. $8.3 \cdot 10^4 \, {\rm yr}$ (regardless of $M_*$) before disappearing. (3) Accretion starts in the Bondi regime and next gets Eddington-limited for $10^{15} \, {\rm g} < M_{\rm PBH} <  7.6 \cdot 10^{14} \, {\epsilon}^{- 1} \, {\rm g}$, i.e. $10^{15} \, {\rm g} < M_{\rm PBH} <  7.6 \cdot 10^{19} \, {\rm g}$. The turnover takes place at time $t_{\rm to}$ such that $M_{\rm PBH}(t_{\rm to}) \simeq 7.6 \cdot 10^{14} \, {\epsilon}^{- 1} \, {\rm g}$ and can be computed from Eq. (\ref{a5bis1}) by setting $M (0) = M_{\rm PBH}$ and $M (t) = M_{\rm PBH}(t_{\rm to})$. We find $t_{\rm to} \simeq 3.3 \cdot 10^{8} \left[ \left(10^{15} \, {\rm g}/ M_{\rm PBH} \right) - 1.3 \, {\epsilon} \right] {\rm yr}$ and consequently $t_{\rm to} < 3.3 \cdot 10^{8} \, {\rm yr}$. Thereafter, accretion becomes Eddington-limited and proceeds much faster, as explained above. 

Our findings set a strong upper bound on the amount of PBHs in the present Universe. To see how this comes about, suppose that every star in the Milky Way disk captures a PBH during its formation process~\cite{prec4}.  We have shown that in such a situation the star gets swallowed by a PBH with $M_{\rm PBH} > 10^{20} \, {\rm g}$ on a time scale much shorter than $t_{\rm H}$~\cite{commentQQ,commentQQQ}. This would not be the case for lighter PBHs, since they have not enough time to engorge the star. Nevertheless, a similar situation would {\it inevitably} occur on a time scale much shorter than $t_{\rm H}$ -- for {\it any} value of $M_{\rm PBH}$ above the evaporation limit -- once the original star has evolved into a white dwarf, owing to the enhanced accretion triggered by the increased density. In order to avoid such a catastrophic conclusion, we have to demand that only a small fraction of disk stars in the Milky Way do capture a PBH. This requirement implies that condition (\ref{a1E}) should be {\it grossly violated}. As we said, no more than $40 \, \%$ of the mass of a GMC goes into stars and it looks reasonable to expect a star-formation efficiency of order $10 \, \%$, in which case the dilution factor is $\delta \simeq 0.1$. Thus, we conclude that only PBHs with $M_{\rm PBH} \gg 10^{25} \, {\rm g}$ can give a substantial contribution to the dark matter in the Universe. More generally, this bound strongly constraints the initial mass function of PBHs and so it helps to discriminate among their various production mechanisms.

Our results also give rise to an exciting possibility concerning stellar-mass black holes. According to the standard theory of stellar evolution, they are the remnant of supernova explosions of stars more massive than about $20 \, M_{\odot}$~\cite{odot} and their mass invariably turns out to exceed $3 \, M_{\odot}$. Therefore, stellar-mass black holes with mass $M < 1 \, M_{\odot}$ can be neither the final stage of stellar evolution nor the product of neutron star-black hole transition through accretion. However, a PBH that swallows a white dwarf ends up with a mass equal to that of the white dwarf itself, namely in the range 
$0.2 \, M_{\odot} - 1.4 \,M_{\odot}$. Thus, observational evidence for black holes with mass $M \lesssim 1 \, M_{\odot}$ would point to the existence of PBHs more massive than $10^{15} \, {\rm g}$.

Obviously, several aspects of the scenario outlined above require further investigation. For instance, it is not clear to us whether an explosive event is produced when the star gets swallowed by a PBH. Still, the characteristic overluminous phase that precedes the disappearance of the star -- whose time span is independent of the stellar mass -- should be actually detectable. 

We thank an anonymous referee for criticism which has led to a substantial improvement of this Letter. One of us (M. R.) would like to thank the Dipartimento di Fisica Nucleare e Teorica, Universit\`a di Pavia, for support.


\begin{thebibliography}{99}
\bibitem{precis2} We note in passing that the numerical coefficient in Eq. (13) of SSQF should read 1.1 rather than 0.51.
\bibitem{comment2} The accretion radius is so small that taking the central values for the density and the temperature is absolutely correct for a PBH in the central region of the star.
\bibitem{prec4} Subsequent PBH capture by a star turns out to be negligible, as pointed out in ref.~\cite{abram}.
\bibitem{commentQQ} This follows from our previous estimate of $t_{\rm to}$ and shows that the PBH capture and its settling at the star centre indeed occur early enough on the cosmic time scale.
\end{thebibliography}
\end{document}